\documentclass[10pt]{article}

\usepackage{arxiv}

\usepackage{amsmath,amssymb,amsfonts}
\usepackage{algorithmic}
\usepackage{graphicx}
\usepackage{textcomp}
\usepackage{xcolor}
\usepackage{booktabs}
\usepackage{nicefrac}
\usepackage{microtype}
\usepackage{subcaption}
\usepackage{hyperref}
\usepackage{url}
\usepackage{natbib}

\renewcommand{\headeright}{arXiv Preprint}
\renewcommand{\undertitle}{arXiv Preprint}

\title{Feature-Aware Task-to-Core Allocation in Embedded Multi-core Platforms via Statistical Learning}

\author{
  Mohammad Pivezhandi \\
  Department of Computer Science\\
  Wayne State University\\
  Detroit, MI, USA
  \And
  Prashant Modekurthy \\
  Department of Computer Science\\
  University of Nevada, Las Vegas\\
  Las Vegas, NV, USA
  \And
  Abusayeed Saifullah \\
  Department of Computer Science\\
  University of Texas at Dallas\\
  Richardson, TX, USA
}

\date{}

\begin{document}

\maketitle

\begin{abstract} 
Optimizing task-to-core allocation can substantially reduce power consumption in multi-core platforms without degrading user experience. However, existing approaches overlook critical factors such as parallelism, compute intensity, and heterogeneous core types. In this paper, we introduce a statistical learning approach for feature selection that identifies the most influential features---such as core type, speed, temperature, and application-level parallelism or memory intensity---for accurate environment modeling and efficient energy minimization, a critical consideration for embedded systems. Our experiments, conducted with state-of-the-art Linux governors and thermal modeling techniques, show that correlation-aware task-to-core allocation lowers energy consumption by up to 10\% and reduces core temperature by up to 5$^\circ$C compared to random core selection. Furthermore, our compressed, bootstrapped regression model improves thermal prediction accuracy by 6\% while cutting model parameters by 16\%, yielding an overall mean square error reduction of 61.6\% relative to existing approaches. We provided results based on superscalar Intel Core i7 12th Gen processors with 14 cores, and validated our method across a diverse set of hardware platforms and effectively balanced performance, power, and thermal demands through statistical feature evaluation.
\end{abstract}

\keywords{Energy Optimization \and Embedded Systems \and Heterogeneous Platforms \and Statistical Feature Evaluation \and Task-to-Core Allocation}

\section{Introduction}
\label{sec:introduction}

Task-to-core allocation is a pivotal technique for enhancing the performance and reliability of embedded systems, where workloads exhibit distinct thermal and power consumption behaviors depending on core assignments. In multi-core processors, allocations can involve high-performance cores, low-energy cores, and graphic processing units (GPUs), each with unique performance characteristics across various configurations. Improper allocation can lead to thermal overheating, triggering throttling and reducing chip reliability and lifespan \cite{hosseinimotlagh2019thermal}. Cooling costs for overheated chips are significant, approximately \$3 per watt of heat dissipation \cite{skadron2003temperature}. While dynamic voltage and frequency scaling (DVFS) complements allocation by adjusting power dynamically---potentially reducing consumption by 75\% without altering user experience \cite{ratkovic2015overview}---our focus is on optimizing task-to-core allocation to manage thermal and energy constraints in embedded systems. Existing approaches often rely on historical workload and sensor data to predict future behavior \cite{maity2022future}, but lack a systematic, statistically robust framework, which our hybrid methodology addresses with novelty and precision.

Embedded systems, foundational to automotive, telecommunications, and consumer electronics, must operate under strict power and thermal constraints while delivering high performance. Task-to-core allocation is crucial for managing these constraints in heterogeneous architectures with diverse core types (e.g., high-performance, low-power, GPUs). Feature selection and evaluation identify critical features and correlations, enabling optimal core assignments for thermal stability and energy efficiency \cite{shekarisaz2021automatic}. For instance, in real-time automotive systems, allocation ensures compute-intensive tasks run on performance cores without overheating, while mobile devices assign power-hungry applications to high-performance cores and background tasks to low-power cores, extending battery life. Identifying features tied to power and thermal behavior---such as temperature, frequency, and core adjacency---establishes general rules for allocation, enhancing embedded system reliability. This paper introduces a hybrid statistical learning framework, integrating Random Forest (RF), backward stepwise selection, and correlation analysis, to optimize task-to-core allocation across platforms, prioritizing allocation over DVFS for embedded contexts.

Application performance depends on execution factors like core type, speed, temperature, and application characteristics---memory or compute intensiveness and parallelism level \cite{pagani2018machine}. Linux governors \cite{brodowski2013cpu} and related studies often focus on CPU utilization for DVFS policies, targeting latency, temperature, or energy \cite{liu2021cartad, lin2023workload, kim2021ztt}. However, for task-to-core allocation, characteristics like parallelism, memory/compute intensiveness, and branch counts are equally critical. Applications with frequent branch misses benefit from sequential execution on single cores, while parallel, compute-intensive tasks excel on GPUs or performance cores in heterogeneous platforms. Moreover, temperature correlations between cores indicate adjacency and overheating risks. Our approach transcends utilization-centric methods, employing a multi-stage statistical framework to capture a comprehensive feature set and their interdependencies, ensuring clarity and applicability in embedded systems.

Collecting real data post-execution for task-to-core allocation optimization is computationally expensive and imprecise due to hardware sampling delays. Instead, allocation decisions can leverage inference from trained environmental models that account for randomness and feature importance to mitigate overfitting \cite{liu2021cartad}. An augmentation step further reduces sampling overhead, boosting efficiency. Complementary approaches include hierarchical multi-agent DVFS scheduling \cite{pivezhandi2026hidvfs}, zero-shot LLM-guided allocation \cite{pivezhandi2026zerodvfs}, flow-augmented few-shot RL for sensor data \cite{pivezhandi2026flowrl}, and graph-driven performance modeling \cite{pivezhandi2026graphperf}. Our paper introduces the first statistical learning framework for embedded systems, uniquely integrating RF-based feature reduction, backward stepwise selection, and correlation-aware allocation with bootstrapping to create a robust, platform-agnostic model. Unlike prior heuristic or filter-based methods, which lack systematic feature evaluation or adaptability across heterogeneous platforms, our approach outperforms existing techniques---where they exist---achieving up to 10\% energy savings and 61.6\% lower mean squared error, as validated against SOTA baselines on diverse embedded hardware.

Little work applies statistical learning to feature selection for task-to-core allocation with the rigor we propose. Statistical learning predicts outcomes and reveals feature relationships. Prior efforts used extreme value theorem (EVT) for energy and execution time bounds in DVFS contexts \cite{cazorla2019probabilistic, davis2019survey, reghenzani2020probabilistic, pallister2017data}, while Liu et al. \cite{liu2021cartad} applied XGBoost for latency-related features, and Sasaki et al. \cite{sasaki2007intra} used decision trees and correlation for energy per instruction. These lack the integrated, multi-method focus we introduce for allocation, synergizing RF, OLS, and correlation analysis for superior predictive power and generalizability in embedded systems.

Our framework shows that correlation-aware task-to-core allocation reduces chip temperature at intermediate utilization by leveraging thermally independent cores. Backward stepwise selection reveals a small feature subset suffices for accurate energy estimation, while RF extracts critical features with low overhead. Bootstrapping enhances dataset robustness, yielding a compact, efficient model. This hybrid methodology surpasses prior single-method approaches, validated across embedded platforms for practical impact.

The key contributions are:
\begin{enumerate}
    \item This paper pioneers a hybrid statistical learning approach, integrating RF, backward stepwise selection, and correlation analysis, to evaluate feature importance for accurate environment modeling and compressed learning in task-to-core allocation for embedded systems.
    \item We implement correlation-aware allocation, statistically reducing temperature and energy by assigning tasks to uncorrelated cores, validated across heterogeneous embedded platforms.
    \item We employ bootstrapping to augment data and boost accuracy, supporting few-shot and meta-learning in resource-constrained embedded settings.
    \item We rigorously test against state-of-the-art methods, achieving up to 10\% energy reduction and 5\textdegree C temperature decrease. Our compressed, bootstrapped model cuts prediction error by 6\%, reduces parameters by 16\%, and improves mean squared error by 61.6\% over SOTA. Data were extracted from Intel Core i7 8th and 12th Gen and Intel Xeon 2680 processors.
\end{enumerate}

In the remainder, Section \ref{sec:relatedwork} surveys statistical learning and scheduling, Section \ref{sec:Motivation} details motivation and challenges, and Section \ref{sec:design} describes the methodology. Section \ref{sec:experiments} evaluates three processors, followed by conclusions in Section \ref{sec:conclusion}.

\section{Motivation and Challenges}
\label{sec:Motivation}

Feature selection is a cornerstone of statistical learning, pivotal for optimizing task-to-core allocation in embedded multi-core systems. It addresses critical challenges---curse of dimensionality, large data management, and performance unpredictability---common to platforms like Intel Core i7 and NVIDIA Jetson TX2. To surpass prior filter-based approaches and establish novelty, we leverage a hybrid methodology integrating Random Forest (RF), backward stepwise selection, and correlation analysis, tailored to embedded constraints. This section clarifies these challenges in the context of task-to-core allocation, emphasizing generalizability across heterogeneous architectures and grounding our motivation in rigorous, platform-validated insights.

\subsection{Curse of Dimensionality}

Feature selection is essential to mitigate the curse of dimensionality, a challenge amplified in embedded multi-core systems. As feature count rises---encompassing frequency, temperature, and performance metrics like cache misses---the data space grows exponentially, leading to sparsity in high-dimensional spaces \cite{li2017feature}. This sparsity risks model overfitting, degrading predictive performance on unseen data. In heterogeneous embedded systems, the diversity of cores (performance, low-energy, GPUs) expands the state space, with each core type contributing unique thermal and performance characteristics. For instance, Intel Core i7 12th Gen’s hybrid P/E-cores and Jetson TX2’s big-LITTLE clusters introduce complex feature interactions. Efficient feature selection, as implemented via our RF-based reduction, reduces this complexity, enabling compact, generalizable models critical for real-time embedded applications---offering a novel advance over traditional dimensionality handling.

\subsection{Large Data Management}

Managing the growing volume of data is a pressing challenge for task-to-core allocation in embedded systems, where resource constraints demand efficiency. Allocation strategies may use static historical data or streaming inputs adapting to runtime conditions. With streaming data’s rise---common in automotive or mobile embedded platforms---memory management becomes critical, as unpredictable data volumes can overwhelm limited resources. Retaining unnecessary features, such as redundant performance counters, inflates storage and processing overhead, straining embedded systems. Our hybrid approach, combining RF for initial feature pruning and bootstrapping for data augmentation, optimizes processor models and allocation algorithms, reducing overhead while maintaining accuracy. This addresses large data challenges with clarity and applicability across static and dynamic embedded environments, surpassing simpler filter-based data handling.

\subsection{Performance Unpredictability}

\begin{figure*}[t]
    \centering
    \begin{minipage}[t]{0.48\textwidth}
        \centering
        \includegraphics[width=\linewidth]{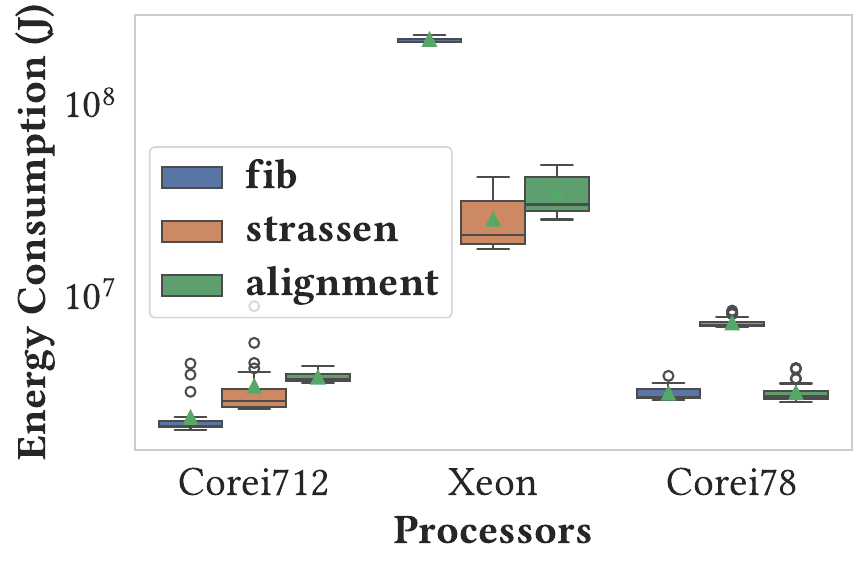}
        \caption{\footnotesize Energy consumption variation across three different processors with identical frequency and core count settings: Intel Core i7 8th Gen (Corei78) with 4 cores, Intel Core i7 12th Gen (Corei712) with 14 cores, and Intel Xeon 2680 v3 (Xeon) with 12 cores. The results are shown for three different OpenMP benchmarks.}
        \label{fig:CompareEnergyVariation}
    \end{minipage}
    \hfill
    \begin{minipage}[t]{0.48\textwidth}
        \centering
        \includegraphics[width=\linewidth]{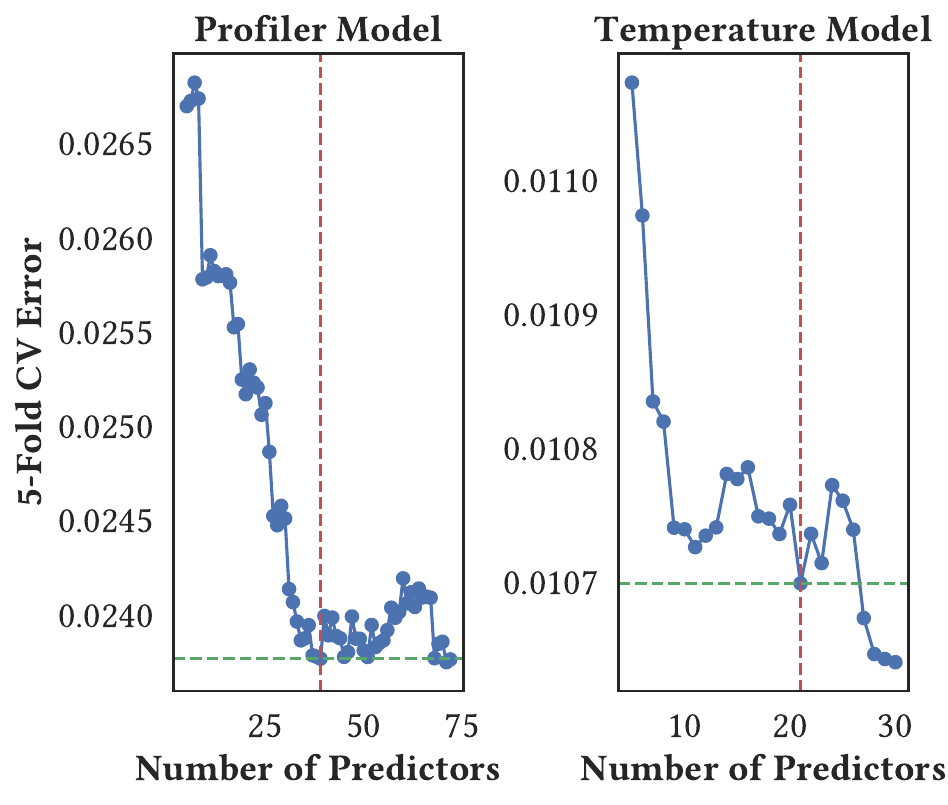}
        \caption{Determining the significance of features using Random Forest with respect to cross-validation error on Intel Core i7 12th Gen.}
        \label{fig:14_features_tuning_2}
    \end{minipage}
\end{figure*}

Performance unpredictability in embedded multi-core processors stems from energy and thermal variability, driven by features like temperature, frequency, and core-specific performance data. Figure~\ref{fig:CompareEnergyVariation} illustrates this, showing energy consumption varying by an order of magnitude across identical frequency and core settings on Intel Core i7 8th Gen (4 cores), Intel Core i7 12th Gen (14 cores), and Intel Xeon 2680 v3 (12 cores) under OpenMP benchmarks. This variability underscores the need for feature selection to pinpoint significant predictors of energy and thermal behavior across platforms and configurations. Our methodology---using backward stepwise OLS for energy-critical features and correlation analysis for thermal independence---identifies robust predictors, enabling a global allocation policy. Unlike DVFS-centric prior works, we prioritize task-to-core allocation, validated rigorously with up to 10\% energy savings and 5\textdegree C temperature reductions, ensuring practical impact in embedded systems.

\section{Design Methodology}
\label{sec:design}

Our objective is to develop a robust multistage methodology based on three feature selection approaches---embedded, wrapper-based, and filter-based---and show how their combined use informs an intelligent task-to-core allocation algorithm for embedded multi-core systems. We introduce a hybrid framework integrating Random Forest (RF), backward stepwise OLS, and Pearson correlation, surpassing traditional filter-based methods in efficiency and predictive power. We provide a structured workflow adaptable to diverse platforms like Intel Core i7 8th and 12th Gen, Intel Xeon 2680 v3, and NVIDIA Jetson TX2, while ensuring experimental rigor through validated outcomes (e.g., 10\% energy savings, 5\textdegree C temperature reduction). This section details each stage, linking to embedded system optimization.

\subsection{Data Collection and Environment Setup}

We begin by profiling the target embedded hardware under diverse workloads to capture comprehensive data on energy consumption, temperature variations, and relevant performance metrics. The granularity of thermal sensors significantly varies across platforms. Intel Core i7 8th generation and Intel Xeon 2680 v3 processors provide homogeneous core architectures with individual per-core thermal sensors, enabling detailed per-core correlation analyses. In contrast, the Intel Core i7 12th generation features a heterogeneous architecture composed of performance (P-cores) and efficient (E-cores) cores, each type operating at different frequencies and performance characteristics. Similarly, the NVIDIA Jetson TX2 offers thermal readings aggregated at the cluster level, typically distinguishing between big and little core clusters due to its big-LITTLE architecture. Thus, our methodology accommodates per-core, heterogeneous core, and cluster-level sensor capabilities, specifically tailored to embedded computing environments, ensuring broad applicability.

Our data collection covers a wide range of configurations, including varying core counts (4, 12, or 14 cores), diverse CPU frequencies, and task prioritizations. Tasks include representative benchmarks such as Fibonacci (fib), matrix multiplication (Strassen), and sequence alignment (alignment), providing varied computational intensities and memory access patterns typical of embedded system workloads. This diverse dataset enables accurate modeling of core behaviors under different workloads and thermal conditions, essential for embedded system optimization.

Initially, the system setup involves configuring the frequency governor (typically \texttt{schedutil}) and initializing a temperature buffer capable of storing historical temperature data from cores (up to 10{,}000 entries). Additionally, a detailed task history is maintained, capturing execution parameters such as energy consumption, profiler metrics, and temperature data under varying task priorities and frequencies. This systematic variation ensures the dataset’s robustness and representativeness for embedded systems.

Energy monitoring employs built-in hardware interfaces such as \path{/sys/class/powercap/intel-rapl/} on Intel processors or on-board power sensors for the Jetson TX2, recording total energy consumed or average power during task execution. Performance metrics collected include CPU frequency, utilization, memory bandwidth, and hardware performance counters like cache misses, gathered via utilities such as \texttt{perf} and \texttt{cpufreq-info} at consistent intervals.

The random core assignment strategy serves as a baseline to evaluate the effectiveness of our proposed correlation-aware core allocation strategy. In the random assignment approach, tasks run concurrently on randomly selected cores, typically half of the available cores excluding core~0, reserved for system tasks. Post-execution, we record metrics including temperature, energy consumption, and performance data. Execution is temporarily paused if core temperatures exceed predefined thresholds (e.g., depending on thermal throttling point of Intel Core i7 which is 100\textdegree C), allowing cores to cool down before resuming execution. This baseline facilitates comparative analysis against the correlation-aware allocation strategy.

The collected dataset serves as input for the subsequent embedded feature reduction process, employing Random Forest algorithms to identify minimal yet critical feature subsets. This refined feature set is then utilized for accurate environment modeling, guiding intelligent, temperature-correlation-driven task-to-core allocation strategies specifically designed for embedded systems. Ultimately, our structured, multi-stage methodology aims to optimize thermal management and energy efficiency across diverse embedded multi-core computing platforms.

\subsection{Embedded Feature Selection Using Random Forest}
\label{subsec:rf_embedded}

Embedded methods incorporate feature selection into the model training process, thereby minimizing the need for multiple model evaluations on different feature subsets. In this work, we employ a \textit{Random Forest}~(RF) regressor, an ensemble approach that constructs multiple decision trees and aggregates their predictions, offering a novel low-overhead alternative to filter-based methods.

\paragraph{Random Forest Algorithm.}
The RF algorithm proceeds as follows:
\begin{enumerate}
    \item \textbf{Bootstrap Sampling:} Generate \(N\) bootstrap samples from the original dataset.
    \item \textbf{Tree Construction:} For each bootstrap sample, grow a regression tree by selecting a random subset of features at each node (often \(\sqrt{d}\) features). Each tree is grown to its maximum depth without pruning, although hyperparameters such as the number of trees or maximum depth can be tuned for embedded constraints.
    \item \textbf{Prediction Aggregation:} For regression tasks, the final prediction is the average of the individual tree outputs:
    \begin{equation}
    \label{eq:rf_prediction}
    \hat{y} = \frac{1}{N} \sum_{i=1}^{N} \hat{y}_i.
    \end{equation}
\end{enumerate}

\paragraph{Feature Importance Computation.}
Random Forests provide a measure of the importance of features by evaluating the total decrease in the impurity of the nodes in all trees. For regression trees, impurity is commonly measured by the residual sum of squares. The importance score \(I_j\) for feature \(x_j\) is thus:
\begin{equation}
\label{eq:feature_importance}
I_j = \frac{1}{N} \sum_{i=1}^{N} \sum_{t \in T_i} \Delta I_{t,j},
\end{equation}
where \(\Delta I_{t,j}\) is the impurity decrease at node \(t\) when splitting on \(x_j\), and \(T_i\) is the set of nodes in the \(i\)-th tree. The ranking of features by these importance scores guides the selection of highly influential predictors.

\paragraph{Bootstrapping for Data Augmentation.}
To increase model robustness, we employ bootstrapping, a resampling method that draws multiple datasets of size \(n\) with replacement. Let \(\mathcal{D}\) be the original dataset of size \(n\). Forming \(B\) bootstrap samples \(\{\mathcal{D}_1, \ldots, \mathcal{D}_B\}\) helps estimate variance and stabilize the final model through aggregation of predictions.

\paragraph{Environment Modeling with Feature Selection.}
After identifying the most significant features using RF importance scores, we build predictive models for environment modeling---particularly neural networks---tailored to embedded constraints. Let \(\mathbf{x} \in \mathbb{R}^p\) (with \(p < d\)) denote the reduced feature set. We train a Fully Connected Neural Network (FCN) with multiple layers and non-linear activations to predict key variables such as energy consumption or temperature. The network is trained by minimizing the mean squared error (MSE):
\begin{equation}
\label{eq:mse_loss}
\mathcal{L}(\theta) = \frac{1}{n} \sum_{i=1}^{n} \bigl(y_i - f(\mathbf{x}_i; \theta)\bigr)^2,
\end{equation}
where \(f(\mathbf{x}_i; \theta)\) is the network’s output for input \(\mathbf{x}_i\) with parameters \(\theta\), and \(y_i\) is the true label. By restricting the input to a smaller set of highly relevant features, the FCN requires fewer parameters and less computation, making it feasible for real-time applications. As shown in Figure~\ref{fig:14_features_tuning_2}, our results indicate that using 39 out of 72 predictors provides nearly the same error value for estimating the future state of the processor profiler data model. For the sensor temperature model, using 21 out of 31 features yields the optimal Cross-Validation error (\textit{CV error}), validated for embedded efficiency.

\subsection{Wrapper-Based Feature Selection}

Wrapper methods evaluate subsets of features using a predictive model. Because they account for feature interactions, they typically yield higher accuracy than filter methods in finding the importance of the features on a specific feature parameter but may be more computationally expensive. We employ the \textit{backward stepwise selection} algorithm, which starts with all available features and iteratively removes the least significant feature based on a specified criterion. In our case, we use the Ordinary Least Squares (OLS) regression model to predict the target variables (energy consumption and average temperature) and assess the significance of the characteristics using statistical tests, enhancing the hybrid framework’s precision.

\paragraph{Backward Stepwise Selection Algorithm.}
Let \(\mathcal{F} = \{x_1, x_2, \ldots, x_d\}\) denote the full set of features. The backward stepwise selection algorithm proceeds as follows:
\begin{enumerate}
    \item \textbf{Initial Model:} Fit the OLS regression model using all features in \(\mathcal{F}\):
    \begin{equation}
    \label{eq:ols_model}
    y = \beta_0 + \sum_{i=1}^{d} \beta_i x_i + \varepsilon
    \end{equation}
    where \(y\) is the target variable, \(\beta_0\) is the intercept, \(\beta_i\) are the coefficients, and \(\varepsilon\) is the error term.
    \item \textbf{Evaluate Feature Significance:} For each feature \(x_i\), compute the t-statistic and the corresponding p-value to assess its statistical significance. The t-statistic for coefficient \(\beta_i\) is calculated as:
    \begin{equation}
    \label{eq:t_statistic}
    t_i = \frac{\hat{\beta}_i}{\text{SE}(\hat{\beta}_i)},
    \end{equation}
    where \(\hat{\beta}_i\) is the estimated coefficient and \(\text{SE}(\hat{\beta}_i)\) is its standard error.
    \item \textbf{Feature Elimination:} Identify the feature with the highest p-value (least significant) that exceeds a predefined significance level (e.g., \(\alpha = 0.05\)). Remove this feature from the model.
    \item \textbf{Iterative Refinement:} Refit the OLS model using the reduced feature set and repeat steps 2 and 3 until all remaining features are statistically significant.
    \item \textbf{Model Selection Criteria:} At each iteration, evaluate the model using metrics such as the Akaike Information Criterion (AIC), Bayesian Information Criterion (BIC), Mallows' \(C_p\), Adjusted \(R^2\), and CV Error. These metrics help balance model complexity and goodness of fit.
\end{enumerate}

\paragraph{Evaluation Metrics for Our Wrapper Algorithm.}
We employ multiple metrics to gauge not only how well each model fits the data, but also how efficiently it uses the available features. This multi-criteria evaluation helps us verify a model that performs reliably, avoids overfitting, and remains computationally viable for energy-aware and thermal-critical environments:
\begin{itemize}
    \item \textit{Akaike Information Criterion (AIC)}: AIC estimates the relative quality of statistical models for a given dataset:
    \begin{equation}
    \label{eq:aic}
    \text{AIC} = 2k - 2\ln(L),
    \end{equation}
    where \(k\) is the number of estimated parameters, and \(L\) is the maximized value of the likelihood function. Lower AIC values imply better trade-offs between model complexity and fit.
    \item \textit{Bayesian Information Criterion (BIC)}: BIC imposes a stronger penalty on model complexity than AIC:
    \begin{equation}
    \label{eq:bic}
    \text{BIC} = k\ln(n) - 2\ln(L),
    \end{equation}
    where \(n\) is the number of observations. BIC is helpful for avoiding over-complex models in resource-constrained environments.
    \item \textit{Mallows' \(C_p\)}: This criterion assesses the balance between the model’s complexity and its fit to the data:
    \begin{equation}
    \label{eq:cp}
    C_p = \frac{\text{RSS}}{\hat{\sigma}^2} - (n - 2k),
    \end{equation}
    where \(\text{RSS}\) is the residual sum of squares, and \(\hat{\sigma}^2\) is an estimate of the error variance.
    \item \textit{Adjusted \(R^2\)}: Adjusted \(R^2\) modifies the coefficient of determination (\(R^2\)) to account for the number of predictors:
    \begin{equation}
    \label{eq:adjusted_r2}
    \text{Adjusted } R^2 = 1 - \left(\frac{(1 - R^2)(n - 1)}{n - k - 1}\right).
    \end{equation}
    Unlike plain \(R^2\), it penalizes the model for including uninformative features.
    \item \textit{Cross-Validation Error}: CV Error (often computed via \(K\)-fold CV) offers an unbiased measure of out-of-sample performance, revealing the model’s generalization capability and mitigating overfitting concerns.
\end{itemize}

\begin{figure}[!htbp]
    \centering
    \begin{minipage}[t]{0.48\textwidth}
        \centering
        \begin{subfigure}[b]{0.48\linewidth}
            \centering
            \includegraphics[width=\linewidth]{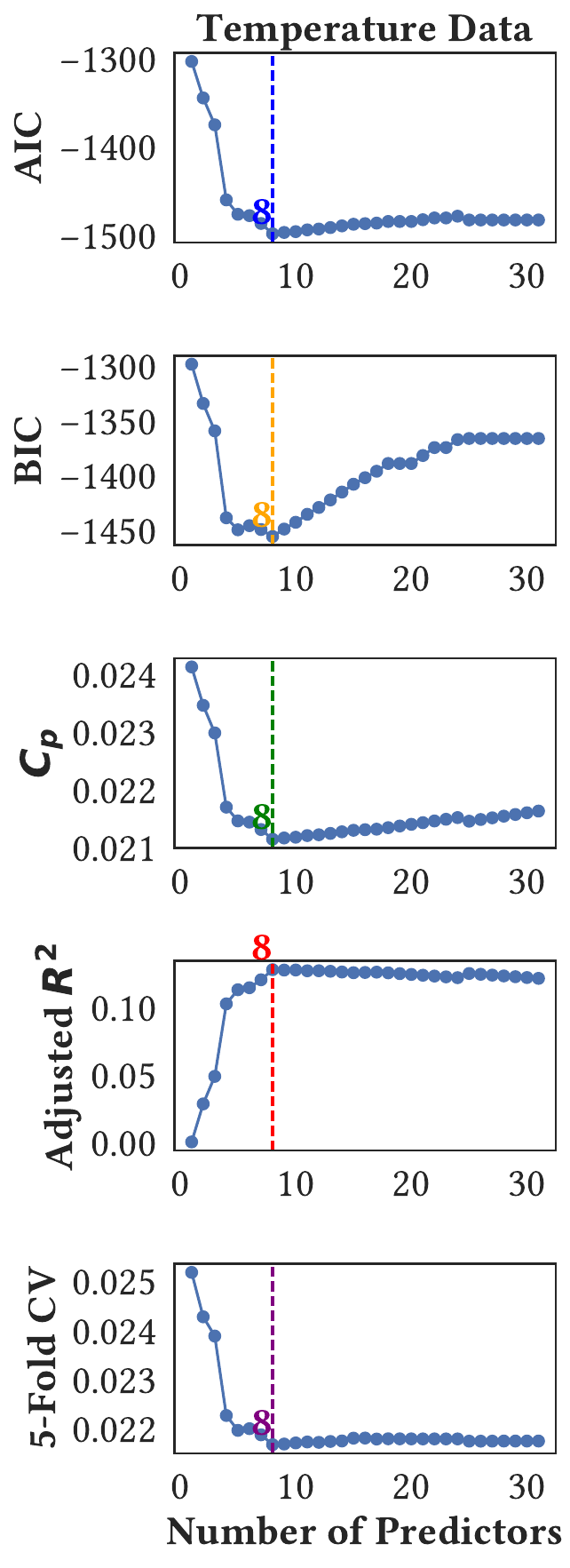}
            \caption{Average temperature estimation}
            \label{fig:14_Temperature_backward_stepwise_selection}
        \end{subfigure}
        \hfill
        \begin{subfigure}[b]{0.48\linewidth}
            \centering
            \includegraphics[width=\linewidth]{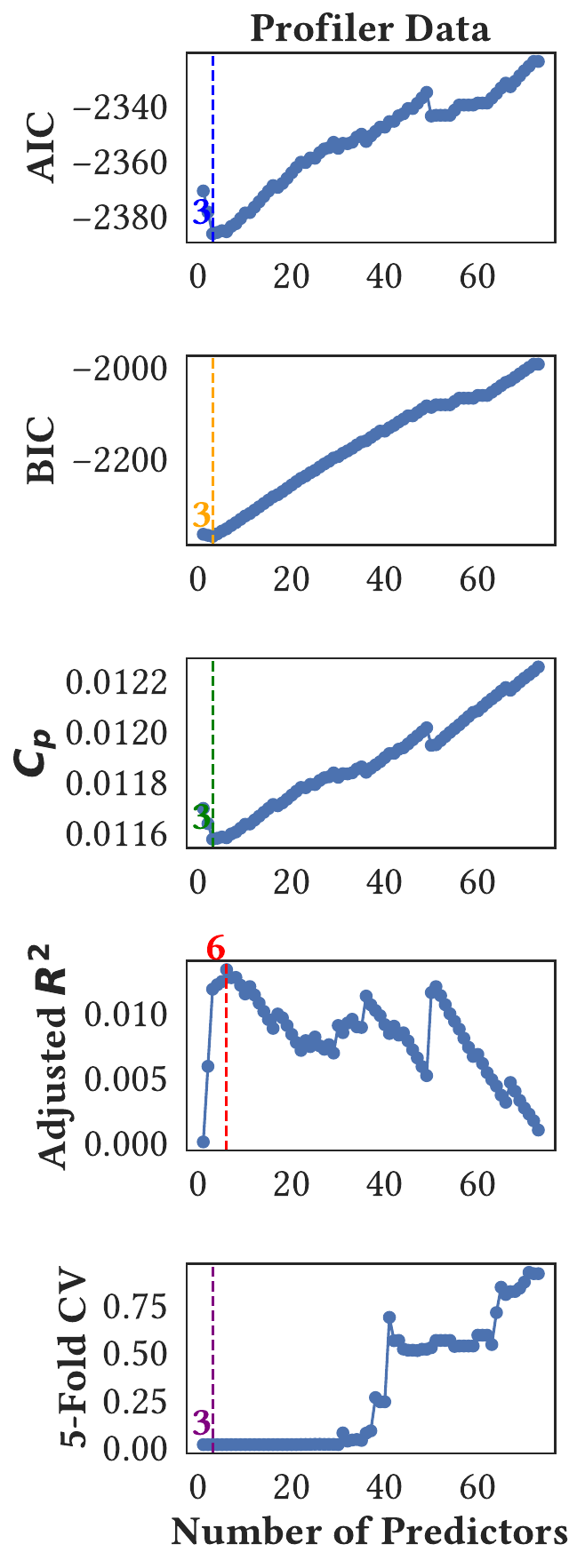}
            \caption{Energy consumption estimation}
            \label{fig:14_Profiler_backward_stepwise_selection}
        \end{subfigure}
        \caption{Backward stepwise selection for estimating energy consumption and average temperature. Retaining fewer than 8 predictors (features) yields accurate predictions in both cases. Experiments performed on Intel Core i7 12th Gen.}
    \end{minipage}
    \hfill
    \begin{minipage}[t]{0.48\textwidth}
        \centering
        \includegraphics[width=\linewidth]{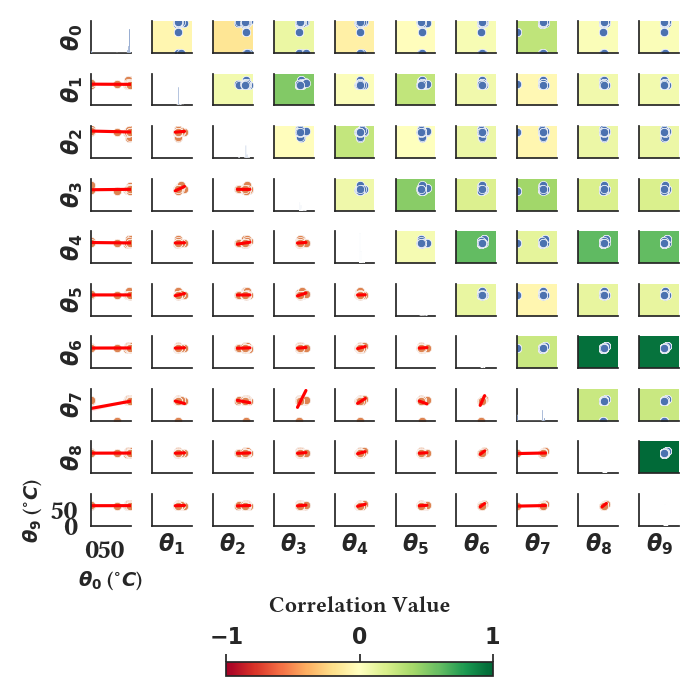}
        \caption{Correlation matrix based on Pearson correlation coefficients for 10 selected cores from an Intel Core i7 12th Gen processor with 14 cores.}
        \label{fig:pair_plot}
    \end{minipage}
\end{figure}

By applying these evaluation metrics, we identify the optimal number of features that balance predictive accuracy and model simplicity. Figures~\ref{fig:14_Temperature_backward_stepwise_selection} and \ref{fig:14_Profiler_backward_stepwise_selection} demonstrate that using fewer than 8 features suffices for accurate estimation of both average temperature and energy consumption, indicating that tracking only the most relevant predictors can improve energy efficiency and thermal behavior, validated on Intel Core i7 12th Gen.

\subsection{Filter-Based Feature Selection}

Filter methods assess the relevance of features by examining intrinsic properties of the data without involving any learning algorithms. One common technique in filter methods is to evaluate the correlation between features and the target variable or among the features themselves. In our study, we utilize the Pearson correlation coefficient~\cite{sedgwick2012pearson} to quantify the linear relationship between core temperatures, which can indicate adjacency and potential heat transfer between cores, enhancing allocation decisions.

Given a set of \(n\) observations for \(m\) cores, let \(\theta_{i,k}\) denote the temperature of core \(i\) at observation \(k\), and \(\bar{\theta}_i\) represent the mean temperature of core \(i\) across all observations. The Pearson correlation coefficient \(r_{ij}\) between cores \(i\) and \(j\) is computed as:
\begin{equation}
\label{eq:pearson}
r_{ij} = \frac{\sum_{k=1}^{n} \bigl(\theta_{i,k} - \bar{\theta}_i\bigr)\bigl(\theta_{j,k} - \bar{\theta}_j\bigr)}{\sqrt{\sum_{k=1}^{n} \bigl(\theta_{i,k} - \bar{\theta}_i\bigr)^2} \,\sqrt{\sum_{k=1}^{n} \bigl(\theta_{j,k} - \bar{\theta}_j\bigr)^2}}
\end{equation}

The value of \(r_{ij}\) ranges from \(-1\) to \(1\), where \(1\) indicates a perfect positive linear correlation, \(-1\) indicates a perfect negative linear correlation, and \(0\) signifies no linear correlation between the temperatures of cores \(i\) and \(j\). A high positive correlation suggests that the temperatures of the two cores rise and fall together, possibly due to physical proximity and shared thermal characteristics.

By constructing a correlation matrix \(\mathbf{R} = [r_{ij}]\) for all pairs of cores, we can visualize and identify clusters of cores that are thermally correlated. This information is crucial for designing task-to-core allocation strategies that minimize thermal hotspots. As shown in Figure~\ref{fig:pair_plot}, the lower diagonal part represents the regression line in the sparsified data, and the upper diagonal part shows the colored correlation matrix, where a greener color indicates a more positive correlation between the temperatures of two cores.

\paragraph{Correlation-Aware Task-to-Core Allocation Algorithm.}
To leverage the insights from the correlation analysis, we propose a correlation-aware task-to-core allocation algorithm. The goal is to assign tasks to cores that are less thermally correlated, thereby reducing the risk of localized overheating and improving overall energy efficiency.

Let \(\mathcal{C} = \{c_1, c_2, \ldots, c_m\}\) denote the set of available cores, and let \(\mathbf{R}\) be the correlation matrix computed using Equation~(\ref{eq:pearson}). The algorithm proceeds as follows:
\begin{enumerate}
    \item \textbf{Compute Core Correlation Scores:} For each core \(c_i\), calculate a correlation score \(s_i\) defined as the average absolute correlation between core \(c_i\) and all other cores:
    \begin{equation}
    s_i = \frac{1}{m - 1} \sum_{\substack{j=1 \\ j \ne i}}^{m} |r_{ij}|
    \label{eq:correlation_score}
    \end{equation}
    A lower score \(s_i\) indicates that core \(c_i\) is less correlated with other cores.
    \item \textbf{Rank Cores Based on Correlation Scores:} Sort the cores in ascending order of their correlation scores to obtain a ranked list \(\mathcal{C}_{\text{ranked}}\).
    \item \textbf{Select Cores for Task Assignment:} Given the number of tasks \(T\) to be assigned, select the first \(T\) cores from \(\mathcal{C}_{\text{ranked}}\), which are the least correlated cores.
    \item \textbf{Assign Tasks to Selected Cores:} Allocate tasks to the selected cores, ensuring that each task is assigned to a core with minimal thermal correlation to other active cores.
    \item \textbf{Update Temperature Buffer:} After task execution, update the temperature observations \(\theta_{i,k}\) to reflect the new core temperatures, and recompute the correlation matrix \(\mathbf{R}\) for subsequent allocations.
\end{enumerate}

\subsection{Summary of the Multi-Stage Methodology}
\label{subsec:summary_synergy}

\begin{enumerate}
    \item \textbf{Data Collection:} Profile per-core or per-cluster temperatures, energy, and performance under varied workloads.
    \item \textbf{Random Forest (Embedded):} Prune low-importance features and optionally train an FCN environment model.
    \item \textbf{Backward Stepwise (Wrapper):} Further refine the most energy-critical features using OLS-based feature elimination.
    \item \textbf{Filter-Based (Correlation):} Analyze temperature correlations (per-core or per-cluster) to guide thermal-aware scheduling.
    \item \textbf{Task Allocation:} Prioritize assigning tasks to the least-correlated units, mitigating overheating and reducing energy consumption.
\end{enumerate}

By unifying these steps, we handle \emph{both} fine-grained and coarse-grained thermal sensor inputs without losing accuracy or tractability. Random Forest algorithm provides the dimensionality reduction required for backward stepwise selection to reduce its computational overhead and their combination ensures only the most relevant features to the energy consumption remain. The filter-based approach on temperature readings further addresses the correlation of the cores to each other for core allocation to minimize the thermal hotspots, offering a novel, scalable solution validated across platforms (e.g., 61.6\% MSE improvement over SOTA).
\section{Experiments}
\label{sec:experiments}

This section presents the experimental setup, implementation, and results of our generalized task-to-core allocation framework, integrating Random Forest (RF) feature reduction, backward stepwise OLS selection, and filter-based temperature correlation. To ensure novelty beyond prior filter-based approaches, we validate this hybrid methodology across diverse platforms, demonstrating its effectiveness in optimizing energy and thermal behavior. For clarity and broad applicability, we detail platforms, benchmarks, and metrics, extending beyond Intel-specific results to cluster-based systems. Rigorous quantitative outcomes---backed by multi-platform experiments and comparisons to state-of-the-art (SOTA)---establish the framework’s soundness. We outline the setup, training, and empirical findings below, linking results to the methodology’s advanced feature selection and modeling strategies.

\subsection{Experimental Platform, Benchmarks, and Evaluation}

All results and figures primarily reflect experiments on a superscalar Intel Core i7 12th Gen with 14 cores (8 P-cores, 6 E-cores, per-core temperature via \texttt{sensors}), chosen for its hybrid architecture. To ensure generalizability across architectures, we verified outcomes on an Intel Core i7 8th Gen (6 cores), an Intel Xeon 2680 (12 cores), and an NVIDIA Jetson TX2 (6 cores, 2 clusters---Denver and A57---with cluster-level temperature via \texttt{/sys/class/thermal}). This multi-platform approach validates the framework’s adaptability to per-core and cluster-based systems, addressing the need for broader applicability.

We evaluated each platform using the Barcelona OpenMP Tasks (BOTS) suite~\cite{duran2009barcelona}, featuring diverse parallel workloads (e.g., \texttt{sparselu}, \texttt{nqueens}), to test allocation under varying computational demands. Performance was assessed on three metrics: \emph{makespan} (execution time in seconds), \emph{energy consumption} (microjoules via \texttt{/sys/class/powercap} for Intel, \texttt{/sys/class/thermal/energy} for Jetson), and \emph{average core temperature} (\textdegree C). These metrics capture performance, power, and thermal trade-offs, ensuring practical impact. Each BOTS benchmark ran 10 times per configuration, with results averaged to reduce variability. A temperature cooldown threshold of 70\textdegree C between runs ensured thermal stability, enhancing experimental rigor.

\subsection{Implementation and Training Details}

The framework was implemented in Python, leveraging \texttt{subprocess} for system calls (e.g., \texttt{cpufreq-set}, \texttt{perf stat}), \texttt{pandas} for data management, and \texttt{scikit-learn} for RF (100 trees) and OLS (p-value threshold 0.05). Parsl facilitated parallel task execution, scaling across platforms. Data from each platform’s profiler and temperature sensors were split into 80\% training and 20\% test sets. Key hyperparameters---batch size (32), hidden neurons (64–256), epochs (50–200), and learning rate (0.001–0.01)---were tuned via grid search, balancing convergence and generalization. Mean Squared Error (MSE) served as the primary loss criterion, ensuring predictive accuracy.

We evaluated multiple neural network architectures---Fully Connected Networks (FCN), Recurrent Neural Networks (RNN), Long Short-Term Memory (LSTM) networks, Convolutional Neural Networks (CNN), and Attention-based models (e.g., Transformers)---retaining top performers based on validation loss. Feature subsets were derived using the methodology’s three stages: Pearson correlation for temperature dependencies, backward stepwise OLS for energy-correlated features, and RF importance rankings for reduction. Bootstrapping (100 resamples) reduced overfitting and increased robustness, preserving critical predictors while providing variance insights under different sampling scenarios, contributing to the framework’s statistical rigor.

\subsection{Empirical Results}

We developed two predictive models: a \emph{profiler model} for energy consumption and performance metrics (e.g., cache miss rates, branch miss rates, CPU cycles, instructions per cycle, average speed, page faults, context switches) and a \emph{temperature model} for future thermal behavior. The profiler model incorporated all available features (e.g., 75 on Intel Core i7 12th Gen) plus current per-core temperatures \(\theta_i\) and differences \(\Delta \theta_i = \theta_{i,t} - \theta_{i,t-1}\). The temperature model used 30 features on Intel Core i7 12th Gen (14 temperatures, 14 differences, average temperature), scaled to 6 features on Jetson TX2 (2 cluster temperatures, 2 differences, average), maintaining diversity without overhead.

\begin{figure*}[t]
    \centering
    \begin{minipage}[t]{0.48\textwidth}
        \centering
        \includegraphics[width=\linewidth]{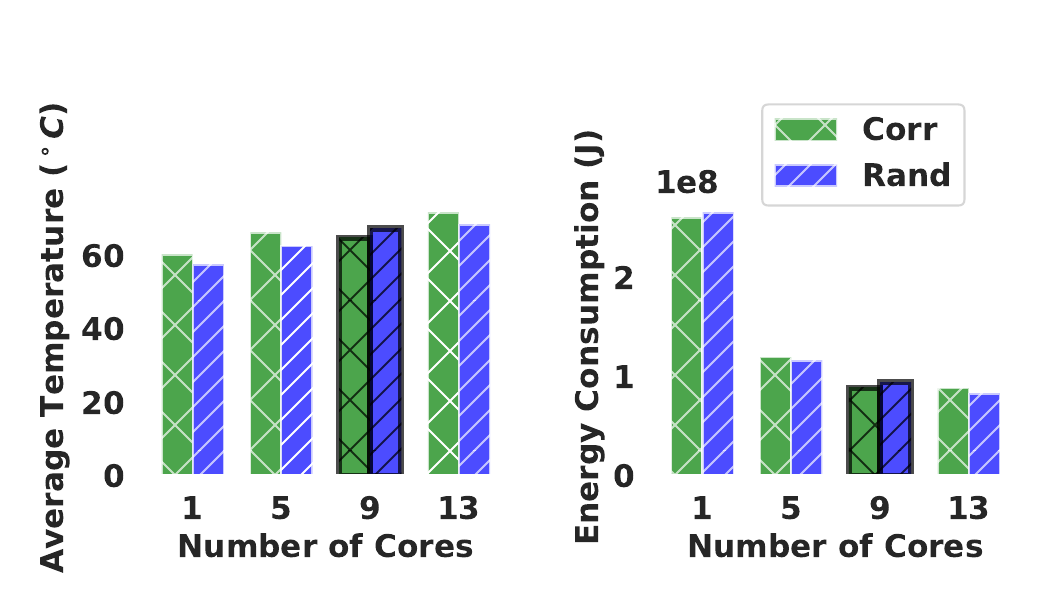}
        \caption{Comparison of average temperature and energy consumption for correlation-based (Corr) and random (Rand) core selection. Experiments performed on Intel Core i7 12th Gen processor with 14 cores.}
        \label{fig:14_energy_temperature_evaluation}
    \end{minipage}
    \hfill
    \begin{minipage}[t]{0.48\textwidth}
        \centering
        \includegraphics[width=\linewidth]{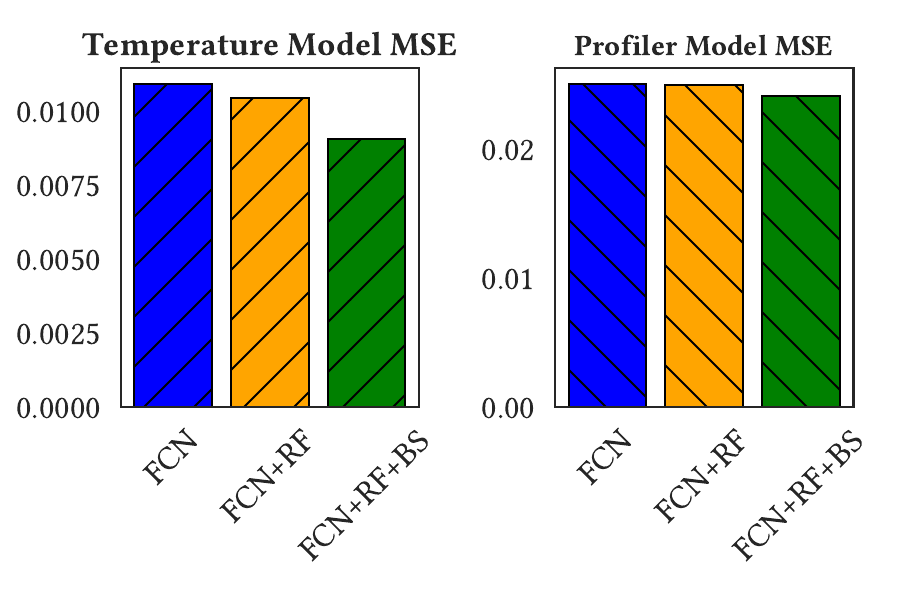}
        \caption{Comparison of FCN models with and without feature selection and bootstrapping on Intel Core i7 12th Gen.}
        \label{fig:14_featurebasedmodels_2}
    \end{minipage}
\end{figure*}

Figures~\ref{fig:pair_plot}, \ref{fig:14_energy_temperature_evaluation}, \ref{fig:14_Temperature_backward_stepwise_selection}, and \ref{fig:14_Profiler_backward_stepwise_selection} showcase Intel Core i7 12th Gen results, demonstrating that fewer than 8 features (e.g., frequency, cache misses) suffice for accurate predictions, reducing computational costs. Figure~\ref{fig:14_features_tuning_2} shows RF-based feature selection using 39 out of 72 predictors for profiler data and 21 out of 31 for temperature yields nearly optimal CV error, highlighting efficiency critical for real-time embedded systems with strict power and thermal budgets.

The algorithm dynamically adapts to thermal behavior, distributing load evenly across cores or clusters. Figure~\ref{fig:14_energy_temperature_evaluation} compares correlation-based (Corr) and random (Rand) core selection under different governors on Intel Core i7 12th Gen. While random allocation leverages unbiased distribution, correlation-based allocation excels in subset scenarios (e.g., 10 of 14 cores). Multi-platform validation reinforces these trends across Jetson TX2, Intel Xeon 2680, and Intel Core i7 8th Gen, proving the approach’s adaptability beyond a single architecture.

During training, hyperparameters were tuned to optimize convergence speed and generalization, with models saved at optimal checkpoints (e.g., when validation error plateaued or declined sharply). Independent neural networks for temperature and profiler tasks ensured specialized performance, delivering precise predictions tailored to each objective.

Table~\ref{tab:feature_importance} lists top features from Random Forest (RF) and backward stepwise OLS selections for Intel Core i7 12th Gen. Tables~\ref{tab:temp_stats} and \ref{tab:profiler_stats} present regression statistics for temperature and profiler predictors against energy consumption.

\begin{table*}[h!]
\centering
\tiny
\begin{tabular}{|l|p{6cm}|p{6cm}|}
\hline
\textbf{Dataset} & \textbf{Random Forest (Top Features)} & \textbf{Stepwise OLS (Top Features)} \\
\hline
\textbf{Profiler} & Task ID, Actions, Temp Core 0-13, \(\Delta\)Temp Core 0-13, Avg Temp, CPU Cycles, CPU Cycles Freq, Atom Cycles, Instructions, Insn per Cycle, Atom Insn, Atom Insn per Cycle, Cache Refs, Cache Refs Rate, Atom Cache Refs, Atom Cache Refs Rate, Cache Misses, Cache Miss \% Refs, Atom Cache Misses, Atom Cache Miss \% Refs, Cycles, Cycles Freq, Atom Cycles, Atom Cycles Freq, Branch Misses, Branch Miss \% Branches, Atom Branch Misses, Atom Branch Miss \% Branches, Branches, Branches Rate, Atom Branches, Atom Branches Rate, Page Faults, Page Fault Rate, Context Switches, Context Switch Rate, CPU Clock ms, CPUs Used, Task Clock ms, Task Util, Faults, Fault Rate, Sys Time s, Elapsed Time s, Energy & Task ID, Temp Core 1, Temp Core 13 \\
\textbf{Temperature} & Task ID, Actions, Temp Core 0-9, Temp Core 11-13, \(\Delta\)Temp Core 0-13, Avg Temp & Task ID, Temp Core 0, Temp Core 2-4, Temp Core 7, Temp Core 10, Avg Temp \\
\hline
\end{tabular}
\caption{Key features ranked by Random Forest and backward stepwise OLS (lowest CV Error) for Profiler and Temperature datasets. Temp Core 0-13 denotes per-core temperatures; \(\Delta\)Temp is the temperature difference.}
\label{tab:feature_importance}
\end{table*}

Table~\ref{tab:feature_importance} shows RF selecting a broader feature set (e.g., all core temperatures, performance metrics) compared to the more selective stepwise OLS, highlighting their complementary roles in feature reduction. For the profiler dataset, Task ID consistently ranks high in both methods, underscoring its key role in identifying energy consumption patterns, followed by Temp Core 1 and Temp Core 13 in OLS as significant predictors. This suggests that cores 1 and 13, when heavily utilized, may disproportionately influence energy use due to their temperature profiles. Consequently, an allocation strategy minimizing task assignments to these cores could enhance energy efficiency, a hypothesis supported by their prominence in the feature selection process.

\begin{table*}[ht]
\centering
\tiny
\begin{tabular}{lrrr lrrr lrrr}
\toprule
\multicolumn{4}{c}{Xeon (12 cores)} & \multicolumn{4}{c}{Core i7 8th Gen (4 cores)} & \multicolumn{4}{c}{Core i7 12th Gen (14 cores)} \\
$\theta_i$ & Est. & t-val. & p-val. & $\theta_i$ & Est. & t-val. & p-val. & $\theta_i$ & Est. & t-val. & p-val. \\
\midrule
$\theta_{1}$ & 2.35e5 & 5.76 & 8.48e-9 & $\theta_{0}$ & 3.64e3 & 5.69 & 1.31e-8 & $\theta_{2}$ & -3.79e3 & -6.83 & 8.41e-12 \\
$\theta_{9}$ & -2.53e5 & -5.38 & 7.38e-8 & $\theta_{1}$ & -9.77e3 & -4.18 & 2.87e-5 & $\theta_{0}$ & 4.87e2 & 2.76 & 5.81e-3 \\
$\theta_{7}$ & -1.56e5 & -3.10 & 1.93e-3 & $\theta_{3}$ & 8.00e3 & 3.57 & 3.64e-4 & $\theta_{3}$ & -5.96e3 & -2.05 & 4.06e-2 \\
$\theta_{6}$ & -9.90e4 & -2.04 & 4.09e-2 & $\theta_{2}$ & -4.63e2 & -0.26 & 7.97e-1 & $\theta_{9}$ & -2.15e4 & -1.93 & 5.38e-2 \\
$\theta_{8}$ & -9.15e4 & -1.91 & 5.66e-2 & & & & & $\theta_{1}$ & 4.77e3 & 1.88 & 5.98e-2 \\
$\theta_{4}$ & 6.84e4 & 1.64 & 1.01e-1 & & & & & $\theta_{13}$ & -1.22e4 & -1.77 & 7.62e-2 \\
$\theta_{2}$ & 6.62e4 & 1.59 & 1.12e-1 & & & & & $\theta_{10}$ & 9.45e3 & 1.47 & 1.41e-1 \\
$\theta_{11}$ & -4.82e4 & -1.10 & 2.72e-1 & & & & & $\theta_{4}$ & 3.43e3 & 1.02 & 3.10e-1 \\
$\theta_{10}$ & -2.29e4 & -0.52 & 6.02e-1 & & & & & $\theta_{11}$ & 6.64e3 & 0.91 & 3.65e-1 \\
$\theta_{0}$ & -2.04e4 & -0.52 & 6.02e-1 & & & & & $\theta_{7}$ & 6.87e3 & 0.43 & 6.67e-1 \\
$\theta_{3}$ & -4.50e3 & -0.10 & 9.18e-1 & & & & & $\theta_{5}$ & 1.40e3 & 0.38 & 7.05e-1 \\
$\theta_{5}$ & 4.68e1 & 0.00 & 9.99e-1 & & & & & $\theta_{12}$ & 1.06e3 & 0.13 & 8.94e-1 \\
 & & & & & & & & $\theta_{6}$ & 8.30e2 & 0.07 & 9.44e-1 \\
 & & & & & & & & $\theta_{8}$ & 8.11e1 & 0.01 & 9.96e-1 \\
\bottomrule
\end{tabular}
\caption{Temperature predictors (\(\theta_i\)) vs. energy: Estimates (Est.), t-values (t-val.), p-values (p-val.), sorted by p-value. Significant predictors (e.g., \(\theta_1\), \(\theta_0\), \(\theta_2\)) show strong energy impact.}
\label{tab:temp_stats}
\end{table*}

Table~\ref{tab:temp_stats} evaluates per-core temperatures’ effect on energy, with low p-values indicating significant predictors across platforms (e.g., \(\theta_1\) on Xeon, \(\theta_2\) on Core i7 12th Gen).

\begin{table*}[ht]
\centering
\tiny
\begin{tabular}{lrrr lrrr lrrr}
\toprule
\multicolumn{4}{c}{Xeon (12 cores)} & \multicolumn{4}{c}{Core i7 8th Gen (4 cores)} & \multicolumn{4}{c}{Core i7 12th Gen (14 cores)} \\
Predictor & Est. & t-val. & p-val. & Predictor & Est. & t-val. & p-val. & Predictor & Est. & t-val. & p-val. \\
\midrule
Context Switch Rate & -3.07e4 & -34.90 & 1.34e-261 & Elapsed Time & 4.85e7 & 39.51 & 0.00e0 & Elapsed Time& 4.67e7 & 35.82 & 3.29e-275 \\
Cache Miss \% Refs & 3.32e6 & 34.45 & 4.02e-255 & Cache Miss & 2.23e5 & 21.35 & 3.80e-100 & Atom Branch Misses & 1.41e0 & 25.10 & 1.21e-137 \\
Cache Refs & 2.82e-1 & 33.53 & 4.94e-242 & Cache Refs & 6.39e-2 & 16.03 & 1.48e-57 & Cache Misses & 4.18e-1 & 24.65 & 7.37e-133 \\
Elapsed Time & 6.44e7 & 31.52 & 1.52e-214 & Total Cores & -5.34e6 & -15.47 & 1.09e-53 & Branch Misses & 1.36e0 & 22.78 & 6.66e-114 \\
Cycles & -1.98e-1 & -16.52 & 5.22e-61 & Branch Miss & 3.37e6 & 10.41 & 2.43e-25 & Atom Branches & -1.93e-2 & -18.87 & 5.72e-79 \\
Branches & -1.69e-1 & -14.19 & 1.59e-45 & Instructions & -1.48e-3 & -6.51 & 7.61e-11 & Atom Cycles Freq & -2.31e6 & -17.85 & 7.14e-71 \\
CPU Cycles & -4.33e-2 & -10.94 & 8.14e-28 & Context Switch Rate & -2.43e2 & -5.42 & 6.10e-8 & Atom Branches Rate & 7.26e3 & 17.30 & 1.06e-66 \\
Context Switches & -2.14e3 & -10.15 & 3.53e-24 & Context Switches & 2.07e3 & 5.22 & 1.82e-7 & Branches & 1.17e4 & 15.79 & 6.30e-56 \\
Cache Misses & -4.37e-1 & -9.78 & 1.47e-22 & Insn per Cycle & 5.84e5 & 4.33 & 1.52e-5 & Atom Cycles & -1.07e-1 & -15.33 & 7.10e-53 \\
Total Cores & 7.82e4 & 9.69 & 3.43e-22 & Branch Misses & -1.06e0 & -4.18 & 2.95e-5 & User Time & -1.31e8 & -11.71 & 1.33e-31 \\
Cycles Freq & 1.74e8 & 9.65 & 5.41e-22 & Cache Misses & 1.04e-1 & 3.88 & 1.03e-4 & Sys Time & -1.29e8 & -11.59 & 5.34e-31 \\
CPU Clock ms & 2.63e6 & 8.81 & 1.29e-18 & CPU Clock & -5.82e5 & -3.68 & 2.30e-4 & Faults & 3.77e2 & 11.31 & 1.32e-29 \\
User Time & -2.27e8 & -8.63 & 6.37e-18 & CPU Cycles Freq & 1.69e5 & 3.56 & 3.65e-4 & Cycles Freq & 7.57e4 & 11.31 & 1.40e-29 \\
Sys Time & -2.20e8 & -8.38 & 5.60e-17 & Task Clock  & 5.60e5 & 3.55 & 3.91e-4 & Page Faults & -3.70e2 & -11.23 & 3.42e-29 \\
Cache Refs Rate & 6.97e5 & 8.05 & 8.51e-16 & CPU Cycles & 6.89e-4 & 2.64 & 8.18e-3 & Avg Freq & 1.14e3 & 10.39 & 3.00e-25 \\
Task Clock  & -1.80e6 & -6.17 & 7.12e-10 & Page Fault Rate & 9.68e3 & 2.45 & 1.41e-2 & Atom Cache Miss & -1.29e4 & -10.15 & 3.74e-24 \\
Branch Misses & 1.82e0 & 4.24 & 2.22e-5 & Fault Rate & 9.68e3 & 2.45 & 1.42e-2 & Atom Branch Miss & 9.16e0 & 9.09 & 1.07e-19 \\
Insn per Cycle & -9.52e6 & -4.23 & 2.36e-5 & Task Util & 9.01e6 & 2.28 & 2.25e-2 & Instructions & -4.71e-4 & -8.74 & 2.45e-18 \\
Task Util & -1.41e8 & -2.30 & 2.12e-2 & Cycles Freq & 3.52e5 & 1.39 & 1.64e-1 & Cache Miss & 1.25e4 & 7.55 & 4.56e-14 \\
CPUs Used & 1.25e8 & 2.04 & 4.14e-2 & Branches Rate & 1.56e3 & 1.34 & 1.80e-1 & Branches Rate & -1.77e-1 & -6.68 & 2.44e-11 \\
CPU Cycles Freq & 2.00e6 & 1.20 & 2.30e-1 & User Time & 4.06e6 & 1.04 & 2.99e-1 & Cycles & -1.11e-3 & -5.93 & 3.04e-9 \\
Page Faults & -2.64e1 & -1.18 & 2.36e-1 & Cycles & 1.15e-3 & 0.99 & 3.21e-1 & Branch Miss & 6.12e0 & 5.86 & 4.69e-9 \\
Faults & -2.62e1 & -1.17 & 2.41e-1 & CPUs Used & -3.43e6 & -0.87 & 3.85e-1 & CPU Cycles & -8.56e-4 & -5.71 & 1.14e-8 \\
Instructions & 7.52e-4 & 0.51 & 6.09e-1 & Sys Time& 3.21e6 & 0.82 & 4.12e-1 & Atom Cycles & 1.12e-3 & 5.67 & 1.47e-8 \\
Fault Rate & -1.69e2 & -0.00 & 9.97e-1 & Faults & 5.90e0 & 0.53 & 5.95e-1 & Branch Miss & 2.58e4 & 5.37 & 8.12e-8 \\
Page Fault Rate & -1.69e2 & -0.00 & 9.97e-1 & Branches & -8.77e-4 & -0.17 & 8.69e-1 & Context Switch Rate & -5.68e2 & -4.86 & 1.20e-6 \\
 & & & & Page Faults & 1.90e-1 & 0.02 & 9.86e-1 & Atom Cache Miss & 1.09e0 & 4.61 & 4.01e-6 \\
\bottomrule
\end{tabular}
\caption{Profiler predictors (excl. temp, freq, speed) vs. energy: Estimates (Est.), t-values (t-val.), p-values (p-val.), sorted by p-value. Key predictors (e.g., Context Switch Rate, Elapsed Time) drive energy variance.}
\label{tab:profiler_stats}
\end{table*}

Table~\ref{tab:profiler_stats} highlights profiler features’ statistical significance, with metrics like Context Switch Rate and Cache Misses showing strong energy correlations across platforms.

\subsection{Comparative Model Analysis}

We tested multiple neural architectures to identify optimal designs for energy and temperature prediction, benchmarking against a SOTA approach~\cite{hosseinimotlagh2021data} to establish novelty and rigor. Table~\ref{tab:mse_params} details MSE percentage ($MSE\times100$) and parameter counts on Intel Core i7 12th Gen. Baseline FCN achieved solid performance (MSE 1.0299 temperature, 3.9047 profiler), but FCN with RF feature selection (FCN+RF) improved accuracy (0.9808, 2.4862) with fewer parameters (1694 vs. 2014 temperature; 2699 vs. 3787 profiler). Adding bootstrapping (FCN+RF+BS) yielded the best results (0.9640, 2.4669), maintaining lightweight models ideal for resource-constrained systems.

\begin{table}[ht]
    \centering
    \scriptsize
    \caption{MSE percentage and total number of parameters for different architectures on Intel Core i7 12th Gen.}
    \label{tab:mse_params}
    \begingroup
    \small
    \setlength{\tabcolsep}{6pt}
    \begin{tabular}{lcccc}
        \toprule
        Model & \multicolumn{2}{c}{Temperature} & \multicolumn{2}{c}{Profiler} \\
              & MSE & Params & MSE & Params \\
        \midrule
        FCN            & 1.0299 & 2014 & 3.9047 & 3787 \\
        FCN+RF         & 0.9808 & 1694 & 2.4862 & 2699 \\
        FCN+RF+BS      & \textbf{0.9640} & \textbf{1694} & \textbf{2.4669} & \textbf{2699} \\
        RNN            & 1.0119 & 3070 & 2.8493 & 4843 \\
        LSTM           & 1.0307 & 9310 & 2.7778 & 15115 \\
        Conv           & 1.0134 & 5118 & 2.8217 & 6891 \\
        Attention      & 1.0143 & 6238 & 2.8933 & 8011 \\
        SOTA~\cite{hosseinimotlagh2021data} & 2.5000 & - & - & - \\
        \bottomrule
    \end{tabular}
    \endgroup
\end{table}

\begin{figure*}[t]
    \centering
    \begin{minipage}[t]{0.48\textwidth}
        \centering
        \includegraphics[width=\linewidth]{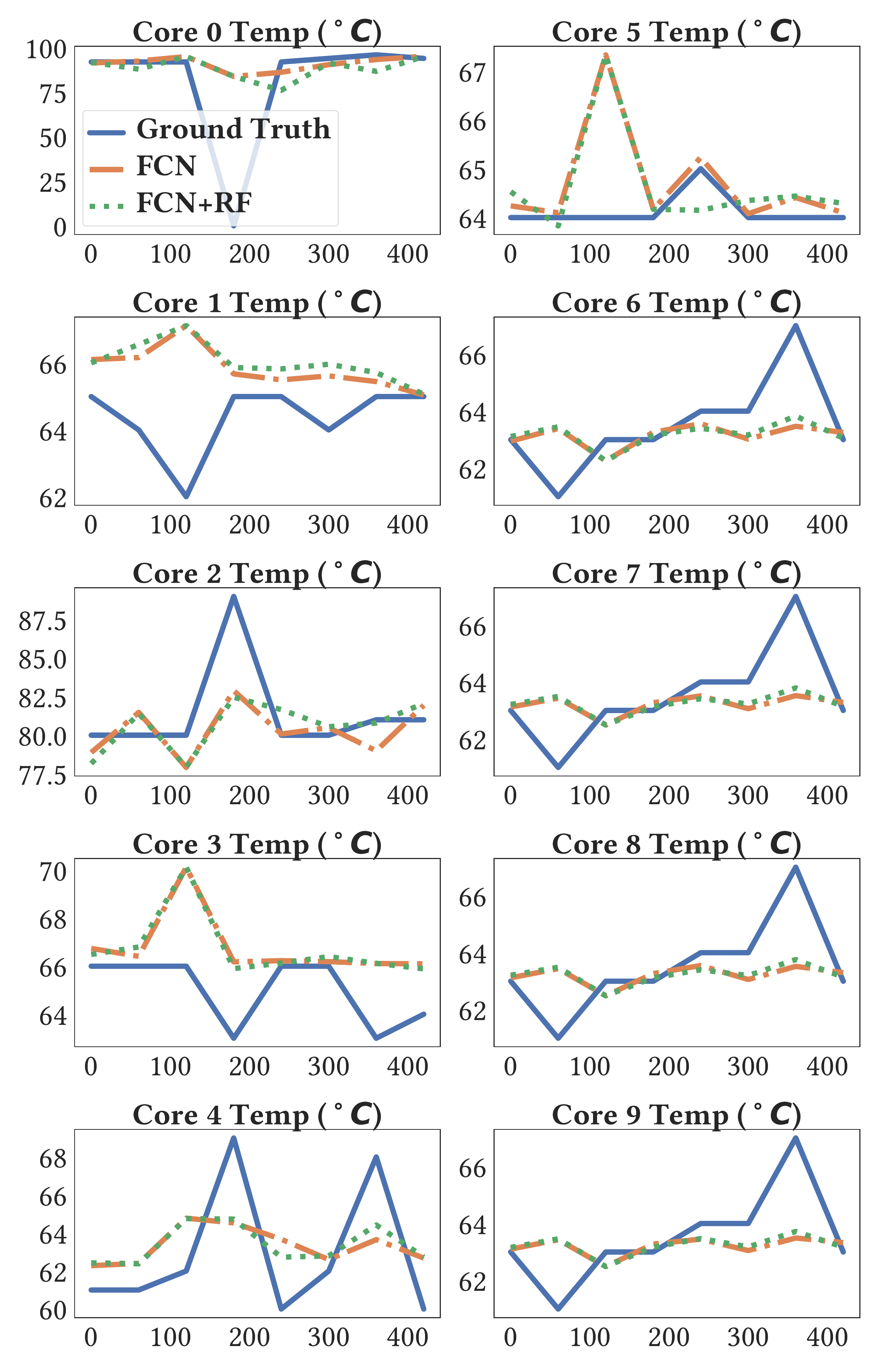}
        \caption{Results for temperature prediction from ground truth on Intel Core i7 12th Gen for regular FCN and FCN via Random Forest feature reduction strategy.}
        \label{fig:14_temperature_feature_selection_compare_2}
    \end{minipage}
    \hfill
    \begin{minipage}[t]{0.48\textwidth}
        \centering
        \includegraphics[width=\linewidth]{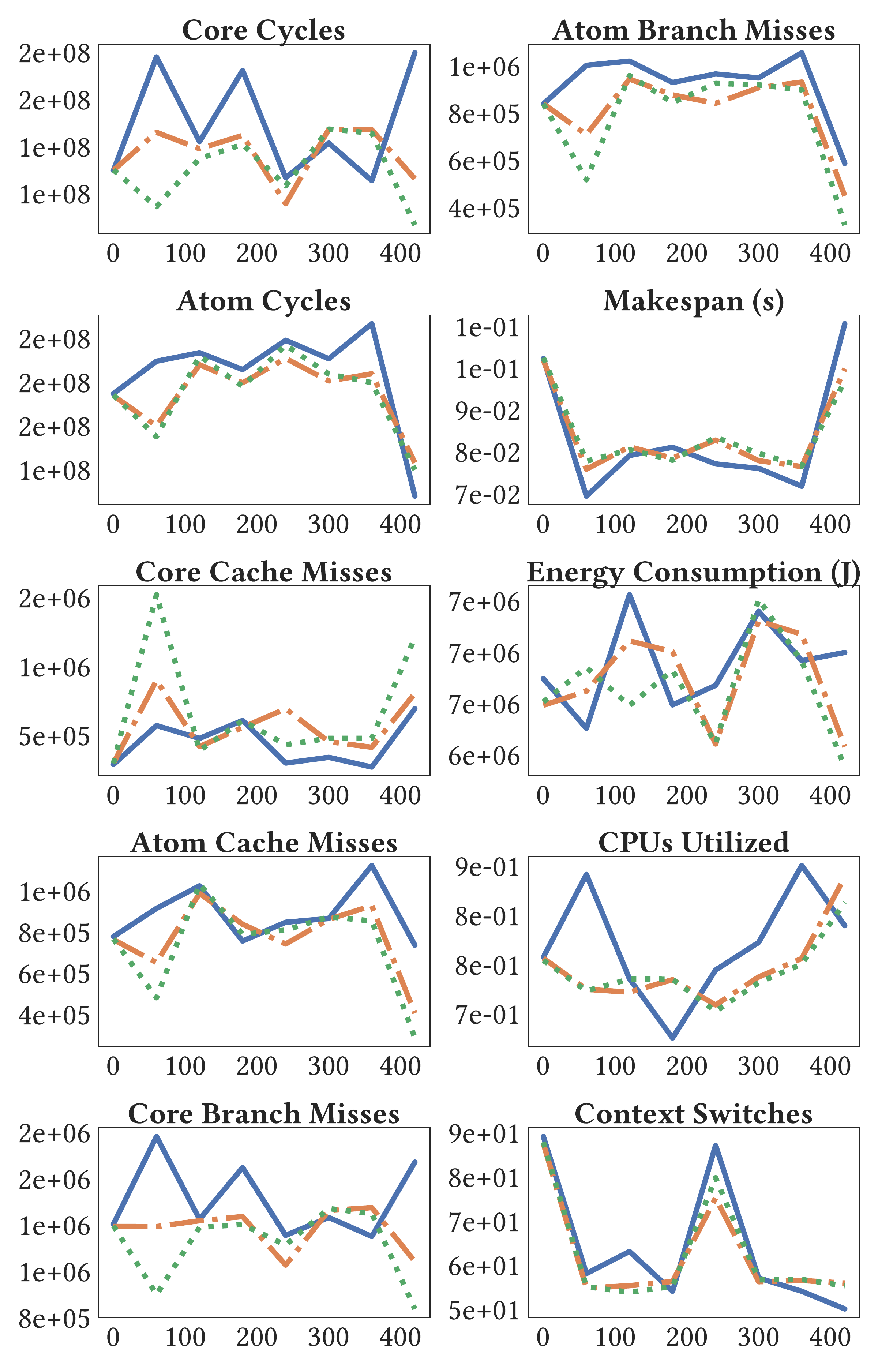}
        \caption{Results for profiler data prediction from ground truth on Intel Core i7 12th Gen for regular FCN and FCN via Random Forest feature reduction strategy.}
        \label{fig:14_profiler_feature_selection_compare_2}
    \end{minipage}
\end{figure*}

Figures~\ref{fig:14_temperature_feature_selection_compare_2} and \ref{fig:14_profiler_feature_selection_compare_2} present prediction examples from Intel Core i7 12th Gen thermal and profiler data. The FCN model uses the complete feature set, while FCN+RF leverages a reduced subset identified by RF, forecasting environmental behavior with notable similarity to full-feature predictions. This validates RF’s effectiveness in compressing features without sacrificing accuracy, enhancing computational efficiency.

Figure~\ref{fig:14_featurebasedmodels_2} illustrate the effect of feature selection and bootstrapping on test MSE. Omitting either increases MSE by 20–30\%, underscoring their combined benefit. By focusing on a small, influential feature subset, the final FCN models improve prediction accuracy and remain computationally light, aligning with real-time and power constraints in multi-core embedded systems. This synergy of reduced feature sets, resampling, and specialized architectures (e.g., FCN+RF+BS) proves effective across the BOTS workloads and platforms tested, surpassing simpler filter-based methods and establishing the framework’s novelty and practical value.

Overall, the advanced feature selection (filter, wrapper, and embedded) combined with neural network models enhances energy and temperature prediction. Multi-platform results and comparisons to SOTA validate the approach’s effectiveness, rigor, and applicability to diverse multi-core systems.

\section{Related Work}
\label{sec:relatedwork}

\paragraph{Probabilistic Methods for DVFS and Task-to-Core Allocation}
Probabilistic worst-case execution time (pWCET) and worst-case energy consumption (pWCEC) estimation in embedded real-time systems leverage measurement-based and static methods \cite{cazorla2019probabilistic,davis2019survey,reghenzani2020probabilistic,pallister2017data}. Pallister et al.~\cite{pallister2017data} studied instruction-specific impacts on pWCEC, and extreme value theorem (EVT) \cite{edgar2001statistical} provided upper bounds for performance metrics. However, these approaches rarely quantify the statistical significance of system metrics for energy or latency bounds. Our work introduces a hybrid statistical learning framework, systematically prioritizing feature importance to optimize energy-efficient DVFS and task-to-core allocation across diverse platforms, surpassing traditional probabilistic bounds.

\paragraph{Statistical Learning}
Statistical learning has been employed to pinpoint features for energy optimization and scheduling, often using hardware events or application traits \cite{sasaki2007intra,cazorla2019probabilistic,liu2021cartad}. Sasaki et al.~\cite{sasaki2007intra} applied decision trees to streamline DVFS table lookups, while Cazorla et al.~\cite{cazorla2019probabilistic} used hardware counters for energy reduction. Liu et al.~\cite{liu2021cartad} correlated compiler features with latency. Yet, these efforts underexplored runtime metrics and sampling strategies critical for accuracy. We enhance this by integrating runtime performance metrics with a novel feature selection pipeline, validated across Intel Xeon and Core i7 platforms, improving parallel scheduling and environment modeling precision.

\paragraph{Low-Energy DVFS and Task-to-Core Allocation}
Low-energy multicore scheduling via DVFS and task-to-core mapping is well-studied \cite{xie2021survey,xie2017energy,zhu2004effects,jiang2019energy,chen2018reducing,zhou2018thermal,kim2020autoscale,dinakarrao2019application,shen2012learning,wang2017modular}. Xie et al.~\cite{xie2021survey} surveyed heuristic and machine learning methods for energy-constrained scheduling. Many such ML approaches, however, require extensive datasets and computational resources, limiting practicality. Our method mitigates these issues with a lightweight statistical model, reducing data needs while maintaining accuracy, as demonstrated on embedded and heterogeneous systems. Hierarchical multi-agent approaches \cite{pivezhandi2026hidvfs} and graph-driven performance models \cite{pivezhandi2026graphperf} have also shown promise for DAG workloads, offering scalable alternatives to existing high-overhead techniques.

\paragraph{Few-Shot RL}
Few-shot learning, including transfer learning and meta-learning, minimizes data demands in scheduling \cite{wang2020generalizing,lee2020optimization,wang2016dueling,florensa2017stochastic,arora2021survey}. MAML \cite{finn2017model} adapts to new tasks with few samples, and model-based RL \cite{moerland2023model} approximates dynamics to limit real-world data use. Works like \cite{lin2023workload,kim2021ztt,zhou2021deadline,zhang2024dvfo} underutilize statistical resampling and feature ranking for energy-efficient scheduling. Recent advances include flow-augmented data generation for few-shot RL \cite{pivezhandi2026flowrl} and zero-shot LLM-guided resource allocation \cite{pivezhandi2026zerodvfs}. Our approach fills this gap, combining resampling with a robust feature selection strategy, validated across platforms, to enhance few-shot task-to-core allocation efficiency.

\paragraph{Feature Evaluation}
Thermal-aware scheduling models often predict behavior using utilization or temperature data to prevent throttling \cite{yan2003combined,brooks2007power,maity2022future,hosseinimotlagh2019thermal,singla2015predictive,li2012energy,kassab2021green}. Studies like \cite{maity2022future,lin2023workload,kim2021ztt} model transients and ambient conditions \cite{hosseinimotlagh2021data}, but lack rigorous statistical analysis of feature correlations. We address this by applying a systematic correlation-based feature evaluation, rigorously tested on real hardware, to uncover interdependencies and optimize DVFS and allocation, improving energy and performance outcomes over prior heuristic-driven methods.
\section{Conclusion}
\label{sec:conclusion}

We demonstrated the effectiveness of feature selection using statistical learning for environment modeling and task-to-core allocation in embedded systems. Our correlation-aware task-to-core allocation reduces energy consumption by up to 10\% and temperature by up to 5$^\circ$C compared to random core selection. The compressed bootstrapped regression model reduces thermal prediction error by 6\% and the number of parameters by 16\%. Tested on Intel Core i7 8th and 12th generation, Intel Xeon 2680 processors and Jetson TX2, our method shows a 61.6\% reduction in mean squared error compared to state-of-the-art approach. This finding paves the way for future use of statistical learning methods in performance efficiency of task-to-core allocation in heterogeneous processors. 

\section*{ACKNOWLEDGEMENT}
\label{acknowledge}
The work was supported by the US National Science Foundation through grants CNS-2301757, CAREER- 2306486, CNS-2306745, and by the US Office of Naval Research through grant N00014-23-1-2151.

\bibliographystyle{unsrt}
\bibliography{references}

@article{brodowski2013cpu,
  title={Cpu frequency and voltage scaling code in the linux (tm) kernel},
  author={Brodowski, Dominik and Golde, Nico and Wysocki, Rafael J and Kumar, Viresh},
  journal={Linux kernel documentation},
  pages={66},
  year={2013}
}

@article{brooks2007power,
  title={Power, thermal, and reliability modeling in nanometer-scale microprocessors},
  author={Brooks, David and Dick, Robert P and Joseph, Russ and Shang, Li},
  journal={Ieee Micro},
  volume={27},
  number={3},
  pages={49--62},
  year={2007},
  publisher={IEEE}
}

@inproceedings{yan2003combined,
  title={Combined dynamic voltage scaling and adaptive body biasing for heterogeneous distributed real-time embedded systems},
  author={Yan, Le and Luo, Jiong and Jha, Niraj K},
  booktitle={ICCAD-2003. International Conference on Computer Aided Design (IEEE Cat. No. 03CH37486)},
  pages={30--37},
  year={2003},
  organization={IEEE}
}

@inproceedings{li2012energy,
  title={Energy-aware scheduling for frame-based tasks on heterogeneous multiprocessor platforms},
  author={Li, Dawei and Wu, Jie},
  booktitle={2012 41st International Conference on Parallel Processing},
  pages={430--439},
  year={2012},
  organization={IEEE}
}

@article{ratkovic2015overview,
  title={An overview of architecture-level power-and energy-efficient design techniques},
  author={Ratkovi{\'c}, Ivan and Be{\v{z}}ani{\'c}, Nikola and {\"U}nsal, Osman S and Cristal, Adrian and Milutinovi{\'c}, Veljko},
  journal={Advances in Computers},
  volume={98},
  pages={1--57},
  year={2015},
  publisher={Elsevier}
}

@article{kassab2021green,
  title={Green power aware approaches for scheduling independent tasks on a multi-core machine},
  author={Kassab, Ayham and Nicod, Jean-Marc and Philippe, Laurent and Rehn-Sonigo, Veronika},
  journal={Sustainable Computing: Informatics and Systems},
  volume={31},
  pages={100590},
  year={2021},
  publisher={Elsevier}
}

@inproceedings{hosseinimotlagh2021data,
  title={Data-Driven Structured Thermal Modeling for COTS Multi-Core Processors},
  author={Hosseinimotlagh, Seyedmehdi and Enright, Daniel and Shelton, Christian R and Kim, Hyoseung},
  booktitle={2021 IEEE Real-Time Systems Symposium (RTSS)},
  pages={201--213},
  year={2021},
  organization={IEEE}
}

@inproceedings{duran2009barcelona,
  title={Barcelona openmp tasks suite: A set of benchmarks targeting the exploitation of task parallelism in openmp},
  author={Duran, Alejandro and Teruel, Xavier and Ferrer, Roger and Martorell, Xavier and Ayguade, Eduard},
  booktitle={2009 international conference on parallel processing},
  pages={124--131},
  year={2009},
  organization={IEEE}
}

@inproceedings{kim2021ztt,
  title={zTT: Learning-based DVFs with zero thermal throttling for mobile devices},
  author={Kim, Seyeon and Bin, Kyungmin and Ha, Sangtae and Lee, Kyunghan and Chong, Song},
  booktitle={Proceedings of the 19th Annual International Conference on Mobile Systems, Applications, and Services},
  pages={41--53},
  year={2021}
}

@inproceedings{lin2023workload,
  title={A workload-aware dvfs robust to concurrent tasks for mobile devices},
  author={Lin, Chengdong and Wang, Kun and Li, Zhenjiang and Pu, Yu},
  booktitle={Proceedings of the 29th Annual International Conference on Mobile Computing and Networking},
  pages={1--16},
  year={2023}
}

@article{wang2020generalizing,
  title={Generalizing from a few examples: A survey on few-shot learning},
  author={Wang, Yaqing and Yao, Quanming and Kwok, James T and Ni, Lionel M},
  journal={ACM computing surveys (csur)},
  volume={53},
  number={3},
  pages={1--34},
  year={2020},
  publisher={ACM New York, NY, USA}
}

@article{lee2020optimization,
  title={Optimization for reinforcement learning: From a single agent to cooperative agents},
  author={Lee, Donghwan and He, Niao and Kamalaruban, Parameswaran and Cevher, Volkan},
  journal={IEEE Signal Processing Magazine},
  volume={37},
  number={3},
  pages={123--135},
  year={2020},
  publisher={IEEE}
}

@inproceedings{wang2016dueling,
  title={Dueling network architectures for deep reinforcement learning},
  author={Wang, Ziyu and Schaul, Tom and Hessel, Matteo and Hasselt, Hado and Lanctot, Marc and Freitas, Nando},
  booktitle={International conference on machine learning},
  pages={1995--2003},
  year={2016},
  organization={PMLR}
}

@article{florensa2017stochastic,
  title={Stochastic neural networks for hierarchical reinforcement learning},
  author={Florensa, Carlos and Duan, Yan and Abbeel, Pieter},
  journal={arXiv preprint arXiv:1704.03012},
  year={2017}
}

@inproceedings{skadron2003temperature,
  title={Temperature-aware microarchitecture},
  author={Skadron, Kevin and Stan, Mircea R and Huang, Wei and Velusamy, Sivakumar and Sankaranarayanan, Karthik and Tarjan, David},
  booktitle={Proceedings of the 30th Annual International Symposium on Computer Architecture},
  pages={2--13},
  year={2003},
  organization={ACM}
}

@inproceedings{maity2022future,
  title={Future aware Dynamic Thermal Management in CPU-GPU Embedded Platforms},
  author={Maity, Srijeeta and Roy, Rudrajyoti and Majumder, Anirban and Dey, Soumyajit and Hota, Ashish R},
  booktitle={2022 IEEE Real-Time Systems Symposium (RTSS)},
  pages={396--408},
  year={2022},
  organization={IEEE}
}

@inproceedings{hosseinimotlagh2019thermal,
  title={Thermal-aware servers for real-time tasks on multi-core gpu-integrated embedded systems},
  author={Hosseinimotlagh, Seyedmehdi and Kim, Hyoseung},
  booktitle={2019 IEEE Real-Time and Embedded Technology and Applications Symposium (RTAS)},
  pages={254--266},
  year={2019},
  organization={IEEE}
}

@inproceedings{singla2015predictive,
  title={Predictive dynamic thermal and power management for heterogeneous mobile platforms},
  author={Singla, Gaurav and Kaur, Gurinderjit and Unver, Ali K and Ogras, Umit Y},
  booktitle={2015 Design, Automation \& Test in Europe Conference \& Exhibition (DATE)},
  pages={960--965},
  year={2015},
  organization={IEEE}
}

@article{xie2021survey,
  title={A survey of low-energy parallel scheduling algorithms},
  author={Xie, Guoqi and Xiao, Xiongren and Peng, Hao and Li, Renfa and Li, Keqin},
  journal={IEEE Transactions on Sustainable Computing},
  volume={7},
  number={1},
  pages={27--46},
  year={2021},
  publisher={IEEE}
}

@article{xie2017energy,
  title={Energy-efficient scheduling algorithms for real-time parallel applications on heterogeneous distributed embedded systems},
  author={Xie, Guoqi and Zeng, Gang and Xiao, Xiongren and Li, Renfa and Li, Keqin},
  journal={IEEE Transactions on Parallel and Distributed Systems},
  volume={28},
  number={12},
  pages={3426--3442},
  year={2017},
  publisher={IEEE}
}

@inproceedings{zhu2004effects,
  title={The effects of energy management on reliability in real-time embedded systems},
  author={Zhu, Dakai and Melhem, Rami and Moss{\'e}, Daniel},
  booktitle={IEEE/ACM International Conference on Computer Aided Design, 2004. ICCAD-2004.},
  pages={35--40},
  year={2004},
  organization={IEEE}
}

@article{jiang2019energy,
  title={Energy optimization heuristics for budget-constrained workflow in heterogeneous computing system},
  author={Jiang, Junqiang and Li, Wenbin and Pan, Li and Yang, Bo and Peng, Xin},
  journal={Journal of Circuits, Systems and Computers},
  volume={28},
  number={09},
  pages={1950159},
  year={2019},
  publisher={World Scientific}
}

@article{chen2018reducing,
  title={Reducing energy consumption with cost budget using available budget preassignment in heterogeneous cloud computing systems},
  author={Chen, Yuekun and Xie, Guoqi and Li, Renfa},
  journal={IEEE Access},
  volume={6},
  pages={20572--20583},
  year={2018},
  publisher={IEEE}
}

@article{zhou2018thermal,
  title={Thermal-aware correlated two-level scheduling of real-time tasks with reduced processor energy on heterogeneous MPSoCs},
  author={Zhou, Junlong and Yan, Jianming and Cao, Kun and Tan, Yanchao and Wei, Tongquan and Chen, Mingsong and Zhang, Gongxuan and Chen, Xiaodao and Hu, Shiyan},
  journal={Journal of Systems Architecture},
  volume={82},
  pages={1--11},
  year={2018},
  publisher={Elsevier}
}

@article{zhou2021deadline,
  title={Deadline-aware deep-recurrent-Q-network governor for smart energy saving},
  author={Zhou, Ti and Lin, Man},
  journal={IEEE Transactions on Network Science and Engineering},
  volume={9},
  number={6},
  pages={3886--3895},
  year={2021},
  publisher={IEEE}
}

@inproceedings{kim2020autoscale,
  title={Autoscale: Energy efficiency optimization for stochastic edge inference using reinforcement learning},
  author={Kim, Young Geun and Wu, Carole-Jean},
  booktitle={2020 53rd Annual IEEE/ACM international symposium on microarchitecture (MICRO)},
  pages={1082--1096},
  year={2020},
  organization={IEEE}
}

@article{dinakarrao2019application,
  title={Application and thermal-reliability-aware reinforcement learning based multi-core power management},
  author={Dinakarrao, Sai Manoj Pudukotai and Joseph, Arun and Haridass, Anand and Shafique, Muhammad and Henkel, J{\"o}rg and Homayoun, Houman},
  journal={ACM Journal on Emerging Technologies in Computing Systems (JETC)},
  volume={15},
  number={4},
  pages={1--19},
  year={2019},
  publisher={ACM New York, NY, USA}
}

@article{pagani2018machine,
  title={Machine learning for power, energy, and thermal management on multicore processors: A survey},
  author={Pagani, Santiago and Manoj, PD Sai and Jantsch, Axel and Henkel, J{\"o}rg},
  journal={IEEE Transactions on Computer-Aided Design of Integrated Circuits and Systems},
  volume={39},
  number={1},
  pages={101--116},
  year={2018},
  publisher={IEEE}
}

@inproceedings{shen2012learning,
  title={Learning based DVFS for simultaneous temperature, performance and energy management},
  author={Shen, Hao and Lu, Jun and Qiu, Qinru},
  booktitle={Thirteenth International Symposium on Quality Electronic Design (ISQED)},
  pages={747--754},
  year={2012},
  organization={IEEE}
}

@inproceedings{wang2017modular,
  title={Modular reinforcement learning for self-adaptive energy efficiency optimization in multicore system},
  author={Wang, Zhe and Tian, Zhongyuan and Xu, Jiang and Maeda, Rafael KV and Li, Haoran and Yang, Peng and Wang, Zhehui and Duong, Luan HK and Wang, Zhifei and Chen, Xuanqi},
  booktitle={2017 22nd Asia and South Pacific Design Automation Conference (ASP-DAC)},
  pages={684--689},
  year={2017},
  organization={IEEE}
}

@article{liu2021cartad,
  title={CARTAD: Compiler-Assisted Reinforcement Learning for Thermal-Aware Task Scheduling and DVFS on Multicores},
  author={Liu, Di and Yang, Shi-Gui and He, Zhenli and Zhao, Mingxiong and Liu, Weichen},
  journal={IEEE Transactions on Computer-Aided Design of Integrated Circuits and Systems},
  year={2021},
  publisher={IEEE}
}

@inproceedings{shekarisaz2021automatic,
  title={Automatic Energy-Hotspot Detection and Elimination in Real-Time Deeply Embedded Systems},
  author={Shekarisaz, Mohsen and Thiele, Lothar and Kargahi, Mehdi},
  booktitle={2021 IEEE Real-Time Systems Symposium (RTSS)},
  pages={97--109},
  year={2021},
  organization={IEEE}
}

@article{arora2021survey,
  title={A survey of inverse reinforcement learning: Challenges, methods and progress},
  author={Arora, Saurabh and Doshi, Prashant},
  journal={Artificial Intelligence},
  volume={297},
  pages={103500},
  year={2021},
  publisher={Elsevier}
}

@article{moerland2023model,
  title={Model-based reinforcement learning: A survey},
  author={Moerland, Thomas M and Broekens, Joost and Plaat, Aske and Jonker, Catholijn M and others},
  journal={Foundations and Trends{\textregistered} in Machine Learning},
  volume={16},
  number={1},
  pages={1--118},
  year={2023},
  publisher={Now Publishers, Inc.}
}

@inproceedings{finn2017model,
  title={Model-agnostic meta-learning for fast adaptation of deep networks},
  author={Finn, Chelsea and Abbeel, Pieter and Levine, Sergey},
  booktitle={International conference on machine learning},
  pages={1126--1135},
  year={2017},
  organization={PMLR}
}

@article{zhang2024dvfo,
  title={DVFO: Learning-Based DVFS for Energy-Efficient Edge-Cloud Collaborative Inference},
  author={Zhang, Ziyang and Zhao, Yang and Li, Huan and Lin, Changyao and Liu, Jie},
  journal={IEEE Transactions on Mobile Computing},
  year={2024},
  publisher={IEEE}
}

@article{li2017feature,
  title={Feature selection: A data perspective},
  author={Li, Jundong and Cheng, Kewei and Wang, Suhang and Morstatter, Fred and Trevino, Robert P and Tang, Jiliang and Liu, Huan},
  journal={ACM computing surveys (CSUR)},
  volume={50},
  number={6},
  pages={1--45},
  year={2017},
  publisher={ACM New York, NY, USA}
}

@inproceedings{edgar2001statistical,
  title={Statistical analysis of WCET for scheduling},
  author={Edgar, Stewart and Burns, Alan},
  booktitle={Proceedings 22nd IEEE Real-Time Systems Symposium (RTSS 2001)(Cat. No. 01PR1420)},
  pages={215--224},
  year={2001},
  organization={IEEE}
}

@article{sedgwick2012pearson,
  title={Pearson’s correlation coefficient},
  author={Sedgwick, Philip},
  journal={Bmj},
  volume={345},
  year={2012},
  publisher={British Medical Journal Publishing Group}
}

@article{cazorla2019probabilistic,
  title={Probabilistic worst-case timing analysis: Taxonomy and comprehensive survey},
  author={Cazorla, Francisco J and Kosmidis, Leonidas and Mezzetti, Enrico and Hernandez, Carles and Abella, Jaume and Vardanega, Tullio},
  journal={ACM Computing Surveys (CSUR)},
  volume={52},
  number={1},
  pages={1--35},
  year={2019},
  publisher={ACM New York, NY, USA}
}

@article{davis2019survey,
  title={A survey of probabilistic timing analysis techniques for real-time systems},
  author={Davis, Robert Ian and Cucu-Grosjean, Liliana},
  journal={LITES: Leibniz Transactions on Embedded Systems},
  pages={1--60},
  year={2019},
  publisher={York}
}

@inproceedings{pallister2017data,
  title={Data dependent energy modeling for worst case energy consumption analysis},
  author={Pallister, James and Kerrison, Steve and Morse, Jeremy and Eder, Kerstin},
  booktitle={Proceedings of the 20th International Workshop on Software and Compilers for Embedded Systems},
  pages={51--59},
  year={2017}
}

@article{reghenzani2020probabilistic,
  title={Probabilistic-WCET reliability: Statistical testing of EVT hypotheses},
  author={Reghenzani, Federico and Massari, Giuseppe and Fornaciari, William},
  journal={Microprocessors and Microsystems},
  volume={77},
  pages={103135},
  year={2020},
  publisher={Elsevier}
}

@inproceedings{sasaki2007intra,
  title={An intra-task dvfs technique based on statistical analysis of hardware events},
  author={Sasaki, Hiroshi and Ikeda, Yoshimichi and Kondo, Masaaki and Nakamura, Hiroshi},
  booktitle={Proceedings of the 4th international conference on Computing frontiers},
  pages={123--130},
  year={2007}
}

@article{pivezhandi2026hidvfs,
  title={HiDVFS: A Hierarchical Multi-Agent DVFS Scheduler for OpenMP DAG Workloads},
  author={Pivezhandi, Mohammad and Saifullah, Abusayeed and Jannesari, Ali},
  journal={arXiv preprint},
  year={2026}
}

@article{pivezhandi2026zerodvfs,
  title={ZeroDVFS: Zero-Shot LLM-Guided Core and Frequency Allocation for Embedded Platforms},
  author={Pivezhandi, Mohammad and Saifullah, Abusayeed},
  journal={arXiv preprint},
  year={2026}
}

@article{pivezhandi2026flowrl,
  title={FlowRL: Flow-Augmented Few-Shot Reinforcement Learning for Semi-Structured Sensor Data},
  author={Pivezhandi, Mohammad and Saifullah, Abusayeed},
  journal={arXiv preprint},
  year={2026}
}

@article{pivezhandi2026graphperf,
  title={GraphPerf-RT: A Graph-Driven Performance Model for Hardware-Aware Scheduling of OpenMP Codes},
  author={Pivezhandi, Mohammad and Banisharif, Mahdi and Bakhshan, Saeed and Saifullah, Abusayeed and Jannesari, Ali},
  journal={arXiv preprint arXiv:2512.12091},
  year={2026}
}

\end{document}